\documentclass[epj]{svjour}

\usepackage{epsfig} 

\usepackage{times}

\usepackage[active]{srcltx}

\renewcommand{\Re}{{\rm Re}}

\renewcommand{\Im}{{\rm Im}}

\begin{document}

%

%\title{\hfill{\tiny FZJ-IKP-TH-2008-XX, HISKP-TH-08/02}\\

%

\title{{\boldmath$A$}-dependence of {\boldmath$\phi$}-meson production in
{\boldmath$p{+}A$} collisions}

\author{A.~Sibirtsev\inst{1,2}, H.-W.~Hammer\inst{1} and 
U.-G.~Mei{\ss}ner\inst{1,3}} 
\institute{
Helmholtz-Institut f\"ur Strahlen- und Kernphysik (Theorie), 
Universit\"at Bonn, Nu\ss allee 14-16, D-53115 Bonn, Germany \and
Excited Baryon Analysis Center (EBAC), Thomas Jefferson National Accelerator
Facility, Newport News, Virginia 23606, USA \and
Institut f\"ur Kernphysik (Theorie), Forschungszentrum J\"ulich,
D-52425 J\"ulich, Germany}

\date{Received: date / Revised version: date}

\abstract{A systematic analysis of the $A$-dependence of $\phi$-meson production
in proton-nucleus collisions is presented. We apply  different formalisms for
the evaluation of the $\phi$-meson distortion in nuclei and discuss the
theoretical uncertainties of the data analysis. The corresponding results are
compared to theoretical predictions. We also discuss the interpretation of the
extracted results with respect to different observables and provide  relations
between frequently used definitions. The perspectives of future
experiments are evaluated and estimates based on our systematical study are 
given.}


\authorrunning{A. Sibirtsev {\it et al.} }
\titlerunning{$A$-dependence of $\phi$-meson production in $pA$ collisions.}
\maketitle
\section{Introduction}

The modification of hadron properties in a nuclear environment remains one of
the most mysterious problems in nuclear   physics, see
e.g.~\cite{Hatsuda1,Bernard1,Bernard2,Brown,Hatsuda2,Li,Saito}.
First experimental results on dilepton spectra from heavy ion collisions
indicated a substantial modification of the spectra in the vicinity of
the $\rho$-meson mass~\cite{Agakishiev1,Masera,Agakishiev2}. Very recent
results from PHENIX for high-energy  $Au{+}Au$ scattering~\cite{Afanasiev} 
also indicated a  significant enhancement of very low mass dielectrons within
the energy range from 150 to 750 MeV. 
This modification of the high mass dilepton spectrum is a key question to
understand the in-medium effects. For instance, the observation of a change of
the $\phi$-meson spectral function would unambiguously clarify the role of
in-medium effects.\footnote{Note that the $\phi$-meson is narrow and well
isolated in mass from other mesons, which makes it a perfect probe of the
in-medium distortion.}
Unfortunately, the statistical accuracy of the recent data~\cite{Afanasiev} is
still not good enough to draw definite conclusions about the modification of the
$\phi$-meson spectral function in heavy ion collisions. 

On the other hand there is a certain belief~\cite{Shor,Adams,Abelev}
that the $\phi$-meson is almost not distorted in nuclear matter and thus can
provide information about  the early partonic stage of heavy ion collisions or
the Quark Gluon Plasma (QGP).  This means that $\phi$-meson production remains
mostly unaffected by hadronic interactions and might serve as the perfect
penetrating probe.  Furthermore, it was proposed~\cite{Shor} that the formation
of the QGP may be detected by enhanced $\phi$-meson production resulting from
the absence of the Okubo-Zweing-Iizuka suppression that is quite substantial in
elementary reactions~\cite{SibirtsevO1,SibirtsevO2,SibirtsevO3}. In this case it
is important that the distortion of the $\phi$-meson is almost  negligible. The
most recent results obtained by the STAR Collaboration~\cite{Abelev} indicate
some distortion of $\phi$-meson production in $Au{+}Au$ collisions, however, it
is difficult to quantify the size of this effect.

Some indications of $\phi$-meson modification at normal nuclear densities have
been reported  from  measurements involving photon~\cite{Ishikawa} and proton
beams~\cite{Abt,Aleev,Daum,Bailey,Tabaru}. The most remarkable observation in
all these measurements is the anomalous $A$-dependence of $\phi$-meson
production. The modification of the spectral function implies the change of the
pole position and the width. When the effective in-medium width of the
$\phi$-meson becomes larger, then its decay probability also increases, leading
to a stronger distortion in nuclear matter. Thus one might
expect~\cite{Oset1,Cabrera,Magas1,Magas2,Hartmann,Sibirtsev1} that the
$A$-dependence indicates a in-medium modification of the imaginary part of the
$\phi$-meson spectral function.

Our previous systematic analysis~\cite{Sibirtsev1} of $\phi$-meson
photoproduction from nuclear targets, however, showed that the observed
$A$-dependence can be well understood by including the $\omega$-$\phi$ mixing in
nuclei. We have investigated single and  coupled-channel phenomena and
reproduced the recent SPring-8 data \cite{Ishikawa} as well as the older Cornell
results at higher energies~\cite{Mcclellan}.

Here we analyze the available data on $\phi$-meson production in proton-nucleus
collisions. Most of the measurements were done at high energies. We provide the
formalism for the data evaluation and give estimates for the theoretical
uncertainties. Moreover, we review different formulas used in  the analysis of
the $A$-dependence at high energies and discuss their compatibility. We also
collect the different theoretical predictions relevant to the $\phi$-meson
distortion in nuclear matter. Finally, we hope that all this information allows
for a systematic understanding of the problem and helps in planning future
strategies for experiment analysis. Therefore, we provide estimates for the data
analysis currently carried out by the ANKE Collaboration at
COSY~\cite{Hartmann}.

\section{Formalism}
\label{formal}

Recently, the eikonal formalism has been considered to extract the in-medium
properties of hadrons from  the $A$-dependence of their production on nuclei
using photon and proton beams. The  distortion of the produced hadrons at low
nuclear matter densities can be expressed in terms of the so-called $t\rho$
approximation. Here $\rho$ is the nuclear density, while $t$ stands for the
forward scattering amplitude. This amplitude is frequently referred to as $f(0)$
and we follow this convention. This amplitude in nuclear matter is not necessary
equal to that in free space because it can be modified by the nuclear
environment. Since this amplitude can be evaluated from the nuclear data, one
might be able to extract the in-medium modification of the hadron interaction.
The $t\rho$ approximation is well applicable at normal, but not high, nuclear
densities and thus can provide a sensitive way for  the nuclear data evaluation.
See, e.g., Refs.~\cite{Magas1,Muhlich1,Muhlich2,Kaskulov} for evaluations using
the $t\rho$ approximation in the eikonal formalism.

Originally this method was proposed to study the interactions of unstable
hadrons with nucleons. Indeed, it is impossible to produce a beam of unstable
hadrons, like $\omega$, $\rho$, $\phi$, {\it etc.} mesons and measure their
scattering on a hydrogen target. However, a nucleus can be used as a source of 
unstable hadrons as well as a target for their interaction and detection.  In
that sense, the eikonal formalism explores the nucleus as a laboratory
constructed at fm distances, {\it i.e.} some kind of \lq\lq Fermilab''. 

Furthermore, the eikonal formalism provides an unique opportunity to evaluate
the scattering length for the interaction of an unstable hadron with a nucleon.
Indeed by measuring the $A$-dependence of the produced particle with different
energies one can extract the scattering amplitude as a function of energy.
Interpolating this amplitude to threshold, one obtains the scattering length.
This scattering length could also be  modified by the nuclear medium and its
extrapolation to free space requires, unfortunately, model dependent
assumptions.

In many applications of the eikonal method the distortion of the produced hadron
is factorized out of the full formalism in order to analyze the data.
Furthermore, the interpretation of the evaluated results is given in terms of
different variables. Thus it is difficult to compare different analyses and to
link the simplified methods that were applied to the original one. For this
reason, we briefly discuss the eikonal formalism for particle production
processes and give the parameter set that can be extracted from the data
analysis. 

\subsection{Eikonal approximation}
\label{sec:1}

The basic idea of the eikonal formalism is to express the interaction of a
particle with a nucleus in terms of effective two-body interactions.  To leading
order, the total reaction amplitude can therefore be built up from a sum of
amplitudes on a single nucleon. The corrections to the leading order come from
multiple interactions. The  eikonal approximation was proposed by
Glauber~\cite{Glauber1,Glauber2} for coherent and incoherent scattering of
hadrons from nuclei. The incident particle is assumed to interact independently
with each target nucleon as it moves along a straight line trajectory through
the nucleus. 

In the original formulation the incident and final particles are identical.
Nevertheless, the formalism can be applied to production processes, {\it i. e.}
when incident and final particles are different. The model was extended by
Formanek and Trefil~\cite{Formanek1,Formanek2} for the case of resonance
production in proton-nucleus collisions and generalized by Berman and
Drell~\cite{Berman1,Berman2}, Ross and Stodolsky~\cite{Ross} and Drell and
Trefil~\cite{Drell1,Drell2} for vector meson photoproduction on nuclei. The best
known extension and application of multi-scattering theory to the coherent and
incoherent production of particles were given by
Margolis~\cite{Margolis1,Margolis2} and K\"olbig and Margolis~\cite{Kolbig}.

In this section, we sketch the formalism of the eikonal approximation necessary
to derive the $A$-dependence of incoherent particle production off nuclei. Our
aim is to illustrate how the elementary amplitude on a quasi-free nucleon is
related to the total production amplitude on a nucleus. This will clarify to
which extent one can extract the elementary amplitude from the experimental
data.  In the following, we will also show how one can interpret this amplitude
with  respect to different nuclear phenomena.

We consider  a two-body interaction at impact vector ${\bf b}$ and describe the
transition from the incident particle $i$ to a final state particle $f$  by the
profile function ${\Gamma}_{if}({\bf b})$ defined as 
\begin{equation}
\Gamma_{if}({\bf b}) = \frac{1}{2\pi i k} \int d^2 q \, 
f_{if}({\bf q}) \, e^{-i{\bf q b}}, 
\label{tran1}
\end{equation}
where $f_{if}$ is the elementary $i{+}N{\to}f{+}X$ transition amplitude at
momentum transfer ${\bf q}$, while $k$ is the projectile particle wave number.
We also define profile functions $\Gamma_{ii}$ and $\Gamma_{ff}$ similar to
Eq.~(\ref{tran1}) with $f_{ii}$ and $f_{ff}$ denoting the amplitudes for 
$i{+}N\to i{+}X$ and $f{+}N{\to}f{+}X$, respectively.

Using  Huygen's principle, the transition amplitude is given by the profile
function as
\begin{equation}
f_{if}({\bf q}) = \frac{ik}{2 \pi} \int d^2b \, \Gamma_{if}({\bf b})
\, \, e^{i{\bf q b}}.
\label{ampl0}
\end{equation}
The distortion of particle $i$ is calculated under the assumption that  each
target nucleon $j$ fixed at transverse (${\bf s}_j$) and longitudinal positions 
($z_j$) independently modifies the  wave of particle $i$  passing through the
target nucleon by the factor\footnote{ This is an essential difference to
cascade-like models that consider distortion on the basis of squared amplitudes.
Our approach takes into account the interference between the distortion
amplitudes and accounts for the quantum dynamics.}
\begin{equation}
1-\Gamma_{ii}({\bf b}-{\bf s}_j).
\end{equation} 
The distortion of the outgoing wave $f$ is defined in a similar way.

The nuclear profile function $\Gamma_{if}^A({\bf b})$ describes the transition
from an initial nuclear state $|I\rangle$ to the final nuclear state $|F\rangle$
as
\begin{eqnarray}
\langle F|\Gamma_{if}^A({\bf b})|I\rangle = \sum_l \,
\langle F| \prod_{z_j<z_l}^{j}\left[1-
\Gamma_{ii}({\bf b}-{\bf s}_j)\right]
\nonumber \\ \times
\Gamma_{if}({\bf b}-{\bf s}_l) 
\prod_{z_m>z_l}^{m}\left[1-
\Gamma_{ff}({\bf b}-{\bf s}_m)\right]\, |I\rangle.
\label{base}
\end{eqnarray}
From this equation it is clear that the nuclear profile function accounts for 
the overall distortion of the incident particle $i$ by nucleons in the target
nucleus before the production of particle $f$  at the positions ${z_j{<}z_l}$,
for the transition $i{+}N{\to}f{+}X$ at $z_l$, and for the distortion of the
produced particle $f$ at ${z_m{>}z_l}$.

Eq.~(\ref{base}) indeed illustrates that the $A$-dependence of particle
production is entirely given by the distortion of both the projectile and the
produced particles. In principle, Eq.~(\ref{base}) summarizes the physical
content of the eikonal approximation.

The  nuclear transition amplitude $f_{if}^A$ is now related to the nuclear
profile function by Eq.~(\ref{ampl0}). The cross section for production of the
particle $f$ in the $i{+}A$ collision, summed over all nuclear final states
$|F\rangle$, is finally given as
\begin{equation}
\frac{d\sigma^A}{dt} = \sum_F |\, f_{if}^A \,|^2,
\label{xsect}
\end{equation}
where $t{=}{-}q^2$ is the four-momentum transfer squared.

The summation over the nuclear final states $|F\rangle$ can be evaluated using
closure and approximating the  many body target wave function $u_I({\bf
r}_1,...,{\bf r}_A)$ ${\equiv}|I\rangle$ by the product of single-particle
density functions as
\begin{equation}
|u_I({\bf r}_1,...,{\bf r}_A)|^2=\prod_{i=1}^{A}\rho_A({\bf r}_i).
\end{equation}

One can assume further that
\begin{eqnarray}
\int \!\!d^2s\,dz\, \Gamma_{ii}({\bf b}{-}{\bf s})\rho ({\bf s},z) 
{\simeq}\int \!\!d^2s \Gamma_{ii}({\bf b}{-}{\bf s}) \nonumber \\
\times \int\limits_{-\infty}^{+\infty}\!\!dz \, \rho ({\bf b},z)=
- f_{ii}(0)\, \frac{2i \pi \, T({\bf b})}{k\, A},
\end{eqnarray}
where  $f_{ii}(0)$ is forward $i{+}N{\to}i{+}N$ scattering amplitude
and the optical thickness function $T({\bf b})$ is defined as
\begin{equation}
T({\bf b}) = A\!\! \int\limits_{-\infty}^{+\infty}\!\!dz \, \rho_A({\bf b},z),
\end{equation}
with  $r^2{=}b^2{+}z^2$. Similar relations also hold for
$f{+}N\to f{+}N$ scattering. Furthermore, we use the relation
\begin{equation}
\left[1+ f_{ii}(0)\, \frac{2i \pi \, T({\bf b})}{k\, A}\right]^A
\simeq \exp\left[\frac{2i \pi \,f_{ii}(0)\, T({\bf b})}{k}\right]
\end{equation}
in order to evaluate Eq.~(\ref{xsect}).

As a consequence, the $A$-dependence of the $f$ particle production in $i{+}A$
collisions becomes a function of the single-particle density function $\rho_A$ 
and of the forward scattering amplitudes for the $i{+}N{\to}i{+}N$ and
$f{+}N{\to}f{+}N$ processes.  The transition amplitude for $i{+}N{\to}f{+}N$
itself does not depend on $A$ under the assumption that it does not depend on
the Fermi motion, which in principle is different for different nuclei. This
assumption is not satisfied for particle production off nuclei at energies below
the reaction threshold in free space, where the Fermi motion is an essential
part of the production process $i{+}N{\to}f{+}N$.

The imaginary part of the forward $f_{ii}$ scattering amplitude is given by the
optical theorem, which is a straightforward consequence of $S$-matrix unitarity
and 
\begin{equation}
{\rm Im} f_{ii}(0)= \frac{k}{4\pi} \sigma_{i},
\label{optic}
\end{equation}
where $\sigma_{i}$ is total cross section for the interaction of the particle
$i$ with a nucleon. Therefore, the forward scattering amplitude can be written
as
\begin{equation}
f_{ii}(0)= \frac{k}{4\pi} \,  (i+\alpha_i)\, \sigma_{i},
\label{optic1}
\label{forward}
\end{equation}
where  $\alpha_i{=}\Re f(0){/}\Im f(0)$ stands for the ratio of the real to the
imaginary part of the forward scattering amplitude. Here one should note that
$\sigma_{i}$ and $\alpha_i$ in nuclear matter are not necessary the same as in a
free space.  Eq.~(\ref{optic}) explicitly illustrates why the total $\sigma_{i}$
reaction cross section is used in the data evaluation.\footnote{Note that
$\sigma_{i}$ can be converted back to the imaginary part of the forward
scattering amplitude or other variables related to the total reaction cross
section, as will be shown below.} A similar relation holds for the final
particle scattering. Since the ratio $\alpha$ is a priori unknown, it is
generally neglected and the differential cross section of Eq.~(\ref{xsect}) is
finally given as
\begin{equation}
\frac{d\sigma^A}{dt} = \frac{d\sigma^N}{dt}
(N_A + \epsilon ),
\label{xsect1}
\end{equation}
where $d\sigma^N{/}dt$ is the elementary cross section for $f$ particle
production in a collision of the particle $i$ with a nucleon, while $N_A$ is the
effective number of target nucleons involved in the interaction,
\begin{equation}
N_A{=}\frac{1}{\sigma_f{-}\sigma_i}
\int\!\!d^2b \,\left[e^{-\sigma_i\, T({\bf b})}{-}
e^{-\sigma_f\, T({\bf b})} \right].
\label{effnum}
\end{equation}

The first term of Eq.~(\ref{xsect1}) is the leading order term, which describes
the production preceded and followed by the distortion of initial and final
particles in the nucleus. Since the production is considered on a single target
nucleon $i{+}N{\to}f{+}X$ this process can be addressed as a direct production
mechanism. If $\sigma_i{=}\sigma_f{=}\sigma$ then  Eq.~(\ref{effnum}) can be
written as
\begin{equation}
N_A = \int\!\!d^2b \, e^{-\sigma T({\bf b})} T({\bf b}),
\label{glaub}
\end{equation}
and for $\sigma{=}0$ we have $N_A{=}A$.  So when initial and final particles are
not distorted by the nucleus, the $A$-dependence of the production process is a
linear function of the atomic mass number. In that case the nucleus is
absolutely transparent, where the nuclear transparency is defined as
\begin{eqnarray}
T_R = N_A/A.
\end{eqnarray}

The second term denoted $\epsilon$ in Eq.~(\ref{xsect1}) contains the
corrections due to the scattering of the  $i$ and $f$ particles out of the
forward direction and contributions from multiple interactions. The
next-to-leading order terms need to be controlled in order to perform the
theoretical evaluation of the data. For this reason, we provide here the results
of numerical calculations which can not be found explicitly in previous studies.

The multiple scattering correction can easily be estimated  under the assumption
that the forward scattering amplitudes $f(q)$ are the same for $i{+}N{\to}i{+}N$
and $f{+}N{\to}f{+}N$ elastic scattering. By parameterizing $f(q)$ as 
\begin{equation}
f(q) =f(0)\exp (-{B \, q^2}),
\end{equation}
with the slope parameter $B$, the correction $\epsilon$ can be obtained as
\begin{equation}
\epsilon =  \sum_{\nu=2}^{A} \frac{1}{\nu} \, \mu^\nu 
\exp[\, B  q^2 (1-\frac{1}{\nu})]\, {\tilde N}_\nu,
\label{epsun}
\end{equation}
where
\begin{equation}
\mu =\frac{1+\alpha^2}{16\pi}\, \frac{\sigma}{B}.
\end{equation}
The multiple scattering collision numbers ${\tilde N}_\nu$ are given by
\begin{equation}
{\tilde N}_\nu = \frac{1}{\nu!}\int d^2b \,\, \sigma^{\nu-1} \,
e^{-\sigma T({\bf b})} \,\, [\, T({\bf b})\, ]^\nu.
\label{multip}
\end{equation}

Furthermore, ${\tilde N}_1$ equals  the effective collision number $N_A$ given
by Eq.~(\ref{glaub}) and represents the leading term of the multiple collision
series.  Additional higher-order corrections given by $\epsilon$ in
Eq.~(\ref{xsect1}) are determined by contributions ${\tilde N}_\nu$ with
$\nu{>}1$. Now $\mu$ can be estimated as the ratio of elastic to the total
interaction cross section and its value can not exceed one.

\begin{table}[t]
\begin{center}
\caption{The multiple scattering collision numbers ${\tilde N}_\nu$ calculated
from Eq.~(\ref{multip}) for a $^{12}C$ target as a function of $\nu$ and total
cross section $\sigma$. The $\nu{=}1$ term is the effective number of target
nucleons involved in the interaction at leading order, while the terms for
$\nu{>}1$ come from the multiple scattering correction $\epsilon$ of
Eq.~(\ref{xsect1}).}
\label{tab1}      
\begin{tabular}{|c|c|c|c|c|c|}
\hline\noalign{\smallskip}
$\sigma$ (mb) & $\nu$=1 & $\nu$=2 &
$\nu$=3 & $\nu$=4 & $\nu$=5\\
\noalign{\smallskip}\hline\noalign{\smallskip}
10 &  8.12 & 1.5 & .25 & .04 & .004 \\
20 &  5.8 & 1.8 & .57 & .16 & .038 \\
30 &  4.3 & 1.7 & .75 & .29 & .102 \\
40 &  3.4 & 1.5 & .79 & .39 & .176 \\
60 &  2.3 & 1.1 & .71 & .46 & .279 \\
80 & 1.8 & 0.8 & .57 & .41 & .301 \\
100 &  1.4 & 0.7 & .46 & .35 & .275 \\
\noalign{\smallskip}\hline\noalign{\smallskip}
\end{tabular}
\end{center}
\end{table}

Note that $\epsilon$ reflects the systematic uncertainty in the application of
the given formalism. To estimate  $\epsilon$ by Eq.~(\ref{epsun}) one needs to
calculate ${\tilde N}_\nu$. The ${\tilde N}_\nu$ terms calculated by
Eq.~(\ref{multip}) for carbon and lead targets are listed in
Tables~\ref{tab1},\ref{tab2} for different cross sections $\sigma$ and $\nu$.
The calculations were performed with a nuclear density function $\rho_A$ taken
as a Wood-Saxon distribution as
\begin{equation}
\rho_A(r)=\frac{\rho_0}{1+exp[(r-R)/d]},
\end{equation}
using the  density parameters~\cite{Knoll}
\begin{equation}
R{=}1.28A^{1/3}{-}0.76{+}0.8A^{-1/3}~\mbox{fm, \,\, }
d{=}\sqrt{3}/\pi~\mbox{fm}. 
\label{radius}
\end{equation}

\begin{table}[b]
\begin{center}
\caption{The multiple scattering collision numbers ${\tilde N}_\nu$ 
calculated using Eq.~(\ref{multip}) for a $^{207}Pb$ target as a function of 
$\nu$ and total cross section $\sigma$. The $\nu{=}1$ term is the first order
effective collision number, while the $\nu{>}1$ terms are due to the multiple
scattering correction $\epsilon$ of Eq.~(\ref{xsect1}).}
\label{tab2}      
\vspace{2mm}
\begin{tabular}{|c|c|c|c|c|c|c|}
\hline\noalign{\smallskip}
$\sigma$ (mb) & $\nu$=1 &$\nu$=2& $\nu$=3 & $\nu$=4 & $\nu$=5
& $\nu$=6 \\
\noalign{\smallskip}\hline\noalign{\smallskip}
10 & 59.9  & 32.3 & 15.0 & 5.8 & 1.9 & 0.5 \\
20 & 24.4  & 17.8 & 13.8 & 9.6 & 5.8 & 3.1 \\
30 & 13.5 & 9.6 & 8.5 & 7.6 & 6.2 & 4.6 \\
40 & 9.1 & 5.9 & 5.2 & 4.9 & 4.7 & 4.3 \\
60 & 5.6 & 3.2 & 2.5 & 2.3 & 2.3 & 2.3 \\
80 & 4.1 & 2.2 & 1.6 & 1.4 & 1.3 & 1.3 \\
100 & 3.2 & 1.7 & 1.2 & 1.0 & 0.9 & 0.8 \\
\noalign{\smallskip}\hline\noalign{\smallskip}
\end{tabular}
\end{center}
\end{table}

The multiple scattering terms given in Tables~\ref{tab1},\ref{tab2} should be
compared to the leading term ${\tilde N}_1{=}N_A$. It is clear that multiple
scattering corrections to the $A$-dependence might be important when $\mu$ is
close to one.  In principle, the ratio $d\sigma^A{/}d\sigma^N$ could be
different from the leading term ${\tilde N}_1{=}N_A$ due to the $\epsilon$
contribution.

For instance, for $pp$ scattering at high energies the elastic cross section
does not exceed $\simeq$15\% of the total cross section, while at COSY energies
this number accounts for $\simeq$40\%. This leads to  large systematical
uncertainties in the evaluation of the $f{+}N{\to}f{+}N$ forward scattering
amplitude from the nuclear data collected at low energies.

Furthermore, ${\tilde N}_\nu$ can not be considered as an effective number of
multi-nucleon clusters in nuclei and should not be  considered as  an estimate
for the production mechanisms involving the interaction of the particles with
few nucleons. The multiple collision numbers ${\tilde N}_\nu$ are attributed to 
elastic scattering of initial ($i$) and final ($f$) particles before and after
the production process $i{+}N{\to}f{+}N$ on a single nucleon.

Finally, within an eikonal approximation, the $A$-dependence for direct particle
production is given by total $i{+}N$ and $f{+}N$  cross sections and by the
single density function $\rho_A$. Since $\sigma_i$ and $\sigma_f$ both depend on
the type of  particle, {\it i.e.} photon, pion, nucleon, etc., as well as on its
kinetic energy, the $A$-dependence also is a function of those degrees of
freedom. The eikonal approximation does not include the $A$-dependence of
particle production due to  the ejectile emission angle.\footnote{The main
advantage the eikonal approximation offers is that the multidimensional
equations reduce to a differential equation in a single variable. This reduction
into a single variable is the result of the straight line approximation
involved.}

\begin{figure}[t]  
\vspace{-3mm}
\centerline{\hspace*{4mm}\psfig{file=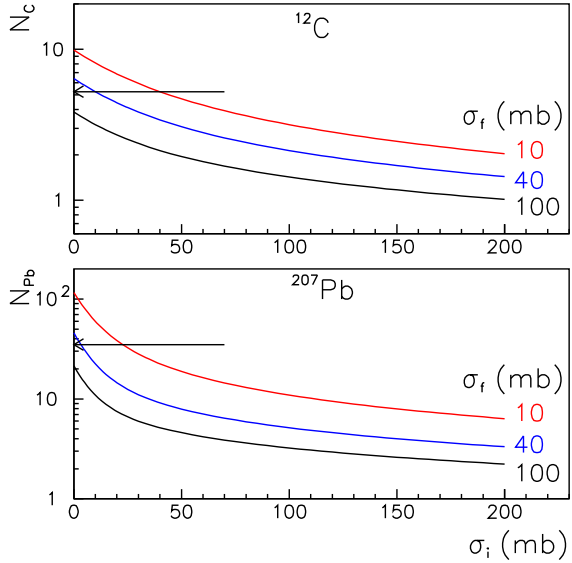,width=8.5cm}}
\vspace*{-3mm}  
\caption{The effective collision numbers $N_A$ calculated by Eq.~(\ref{effnum})
for $C$ and $Pb$ nuclei and for different $\sigma_f$ total cross sections as a
function of $\sigma_i$. The arrows indicate the $A^{2/3}$-dependence.}  
\label{neff5d}  
\end{figure}

Now, the solid lines in Fig.~\ref{neff5d} show the effective collision numbers
$N_A$ calculated by Eq.~(\ref{effnum}) for carbon and lead nuclei  as a function
of $\sigma_i$. The results are shown for $\sigma_f$=10, 40 and 100~mb. The
calculations indicate reasonable sensitivity to both $\sigma_i$ and  $\sigma_f$.

The arrows in Fig.~\ref{neff5d} indicate the $A^{2/3}$ dependence that is
frequently assumed for inelastic hadron-nucleus reactions. The $A^{2/3}$
dependence is in general discussed in terms of the absorptive interaction of the
incident particle at the nuclear surface. However, this not an unique
explanation as is illustrated by Fig.~\ref{neff5d}. The $A^{2/3}$ dependence
might result from various combinations of $\sigma_i$ and $\sigma_f$, which
reflect quite  different physics.

The data analysis of the $A$-dependence of particle production from nuclei is
frequently done in terms of an exponent $\alpha$ fitted to experimental results
for a double differential cross section as
\begin{equation}
\frac{d^2\sigma^A}{d\Omega dT} = c A^\alpha,
\label{slope}
\end{equation}
where $c$ is some constant. In general, the exponent $\alpha$  is evaluated from
data collected at different kinematical conditions, such as emission angles
$\Omega$, kinetic energies $T$ of produced  particles, incident beam energies
and different kinds of projectile ($i$) and ejectile ($f$) particles.  In the
eikonal approximation, the variety of kinematical conditions can easily be
classified by considering the $A$-dependence as a function of the total
$\sigma_i$ and $\sigma_f$  cross sections. However, one should not expect
validity of the eikonal approximation at production angles away from the forward
direction.

Furthermore, the data analysis in terms of the  $A^\alpha$ function introduces
additional systematical uncertainties, which can be well understood by
inspecting Fig.~\ref{neff5d}. We consider the following example: For
$\sigma_f$=10 mb we obtain $\alpha{=}2/3$, as is indicated by the arrows in
Fig.~\ref{neff5d} after fitting both $^{12}C$ and $^{207}Pb$ data. This
$A$-dependence corresponds to $\sigma_i{\simeq}$39 mb for $^{12}C$, as is shown
by the arrow in the upper panel of Fig.~\ref{neff5d}. At the same time the lower
panel for a $^{207}Pb$ target shows that $\alpha{=}2/3$ corresponds to
$\sigma_i{\simeq}$22 mb. This large uncertainty in $\sigma_i$  is reflected in
the standard deviation of the $\alpha$ slope and vice versa. To avoid this
uncertainty in the theoretical analysis it is more useful to analyze the ratio
of the production cross section measured with different targets. 

Nevertheless, for completeness we evaluate
Eq.~(\ref{effnum}) as a function of both $\sigma_i$ and $\sigma_f$ for various
nuclei and fit the calculations by $A^\alpha$ in order to determine the {\em
average} slope $\alpha$. This {\em average} slope $\alpha$ is shown in
Table~\ref{tab3} as a function of $\sigma_i$ and $\sigma_f$. The calculations
were done for $C$,  $Al$, $Cu$,  $Ag$, $Au$, and $Pb$ nuclei. (We do not
indicate the  calculations for $\sigma_i$=$\sigma_f$=0 that obviously result in
$\alpha$=1.) The {\em average} slope parameters $\alpha$ are afflicted with
large uncertainties, e.g. $\alpha=0.35\pm 0.18$ for $\sigma_i = 40$ mb and
$\sigma_f = 30$ mb. For this reason, we discourage fitting cross sections for a
single nucleus and recommend to analyze cross section ratios instead.

The eikonal approximation provides the $A^\alpha$-dependence with
$1{\le}\alpha{\le}0.3$. As is shown in Table~\ref{tab3}, the $A^{2/3}$
dependence can be observed under various conditions given by $\sigma_i$ and
$\sigma_f$. For instance, an $A^{2/3}$ dependence is expected for $\sigma_i$=0
and $\sigma_f$=50~mb, which might correspond to the photoproduction of $\pi$,
$\rho$, $\omega$ and other mesons. Finally, the $A^{2/3}$-dependence can not be
addressed as only due to the  interaction of the incident particle at the
nuclear surface. It is also clear that an interpretation of the $A^{1/3}$
dependence could not be given in a unambiguous way.

\begin{table}[t]
\begin{center}
\caption{
The {\em average} slope $\alpha$ of the $A^\alpha$-dependence fitted to the
effective collision numbers calculated from Eq.~(\ref{effnum}) for different
cross sections $\sigma_i$ and $\sigma_f$. The calculations were done for $C$, 
$Al$,  $Cu$,  $Ag$, $Au$, and $Pb$ nuclei.}
\label{tab3}      
\begin{tabular}{|c|c|c|c|c|c|c|c|c|c|}
\hline\noalign{\smallskip}
$\sigma_f$ & \multicolumn{9}{|c|}{$\sigma_i$ (mb)} \\
\noalign{\smallskip}\hline\noalign{\smallskip}
(mb)&  0 & 10 & 20 & 30 & 40 & 50 & 60 & 70 & 80 \\ 
\noalign{\smallskip}\hline\noalign{\smallskip}
10 & .85 & .66 & .56 & .51 & .48 & .46 & .45 & .44 & .43  \\
20 & .76 & .56 & .46 & .41 & .39 & .37 & .36 & .35 & .35 \\
30 & .71 & .51 & .41 & .37 & .35 & .33 & .33 & .32 & .32  \\
40 & .68 & .48 & .39 & .35 & .33 & .32 & .31 & .31 & .31  \\
50 & .66 & .46 & .37 & .34 & .32 & .31 & .31 & .30 & .30 \\
60 & .65 & .45 & .36 & .33 & .31 & .31 & .30 & .30 & .30 \\
70 & .63 & .44 & .36 & .32 & .31 & .30 & .30 & .30 & .30 \\
80 & .63 & .43 & .35 & .32 & .31 & .30 & .30 & .30 & .30 \\
90 & .62 & .42 & .35 & .32 & .31 & .30 & .30 & .30 & .30 \\
\noalign{\smallskip}\hline\noalign{\smallskip}
\end{tabular}
\end{center}
\end{table}

\subsection{Angular dependence}
Since in the actual experiments the data are collected at some fixed angles or
integrated over a certain angular interval it is important to estimate how much
the $A$-dependence is effected by the ejectile angle. Such an estimate can be
done using a quasi-classical approximation as is shown below.

\begin{figure}[b]  
\vspace{-4mm}
\centerline{\hspace*{4mm}\psfig{file=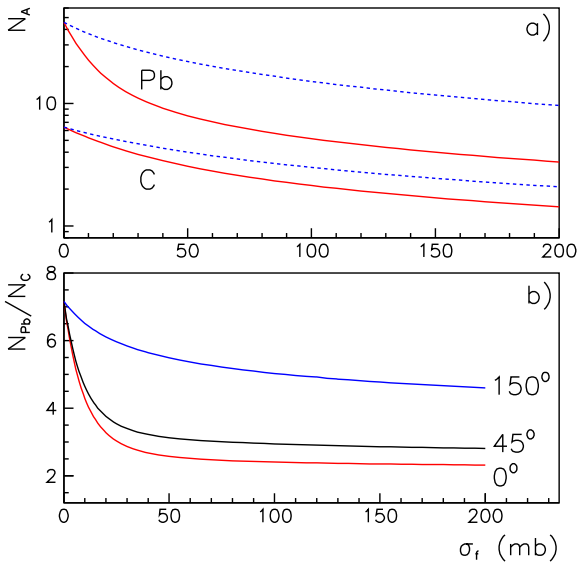,width=8.5cm}}
\vspace{-3mm}  
\caption{a) The effective collision numbers calculated using Eq.~(\ref{effnum1})
for $C$ and $Pb$ nuclei and $\sigma_i$=40~mb as a function of $\sigma_f$. The
solid lines show the results for emission angle $\theta$=0$^o$, while the dashed
lines are the calculations for $\theta$=150$^o$. b) The ratio $N_{Pb}{/}N_C$ as
a function of $\sigma_f$ calculated for angles $\theta$=0$^o$, 45$^o$ and
150$^o$.}  
\label{neff5c}  
\end{figure}  

As a beam of particles $i$ passes through the nucleus, its intensity is
attenuated due to the scattering out of the beam direction. Since particles can
be removed from the beam because of both elastic and inelastic interactions with
the target nucleons, the attenuation is determined by the distortion cross
section $\sigma_i$. 

The attenuation probability of an $i$ particle passing  through the nucleus at
impact parameter ${\bf b}$ and longitudinal positions from $-\infty$ to $z$ is
then given by
\begin{equation}
S_i({\bf b},z){=}\exp\left[{-}
\sigma_i\!\!\int\limits_{-\infty}^z \!\!dz'\,
\rho_A({\bf b},z')\right]{=}e^{-\sigma_i T_z({\bf b})},
\label{class1}
\end{equation}
and $T_z({\bf b})$ can be considered as the linear nuclear density.

Thus Eq.~(\ref{class1}) is a semi-classical description of particle $i$
attenuation in matter and  one might argue that in this case $\sigma_i$ should
be taken as an inelastic or absorption cross section rather than  total reaction
cross section. In that sense an emission of a particle out of the beam
trajectory can be also considered a distortion. That is why $\sigma_i$ is
discussed  as a distortion cross section.

The incident particle $i$ interacts with target nucleon at a transverse  ${\bf
b}$ and longitudinal $z$ coordinate and  produces the final particle $f$. Let us
to consider that $f$ is moving along the line fixed at an azimuthal angle $\phi$
and polar angle $\theta$ with respect to the incident particle beam direction.
The attenuation probability of passing $f$ through the nucleus in that case is
given as 
\begin{equation}
S_f({\bf b},z,\theta,\phi) = \frac{1}{2\pi}
\exp\left[-\sigma_f\!\!
\oint d\xi \ {\rho}(|{\bf r}_{\xi}|)
\right],
\label{sf}
\end{equation}
where the integration is performed  along the path of the produced  particle $f$
defined by \begin{equation} r_{\xi}^2 = (b+ \xi \cos\phi \sin\theta)^2+(\xi
\sin\phi \sin\theta )^2 +(z + \xi \cos\theta )^2.
\end{equation} 
Here we again assume that both initial and final particle move along the
straight trajectories before and after the production vertex, but now the
ejectile trajectory depends on $f$ emission angles $\phi$ and $\theta$. 

Finally, the effective collision number can be evaluated by integration over the
nuclear volume as
\begin{equation}
N_A {=}\!\!\int\! d^2b \, dz\, \rho({\bf b},z)
\,S_i({\bf b},z) \,S_f({\bf b},z,\theta,\phi ),
\label{effnum1}
\end{equation}
and now depends on the production angles. It is easy to show that
Eq.~(\ref{effnum1}) reduces to the eikonal formalism given by Eq.~(\ref{effnum})
after an integration over the azimuthal angle $\phi$ at polar angle
$\theta{=}0^o$.

Now Fig.~\ref{neff5c}a) shows the effective collision numbers calculated from
Eq.~(\ref{effnum1}) for $C$ and $Pb$ nuclei and $\sigma_i$=40~mb as a function
of $\sigma_f$. Here the solid lines indicate the results obtained for a final
particle emission angle $\theta$=0$^o$, while the dashed lines show the
calculations for $\theta$=150$^o$. 

\begin{figure}[t]
\vspace*{-3mm}
\centerline{\hspace*{1mm}\psfig{file=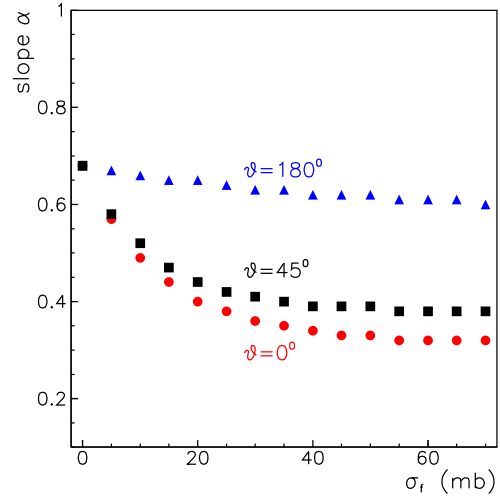,width=7.5cm}}
\vspace*{-4mm}
\caption{The slope of the $A^\alpha$-dependence as a function of the effective
cross section $\sigma_f$ shown for the different production angles:
$\theta{=}0^0$ (circles), $\theta{=}45^0$ (squares) and $\theta{=}180^0$
(triangles). The results are given for the direct production mechanism.}
\label{cofi10}
\end{figure}

The calculations indicate  a quite strong angular dependence of $N_A$.
Fig.~\ref{neff5c}b) shows the ratio of the effective collision numbers
$N_{Pb}{/}N_C$ as a function of $\sigma_f$ for different angles: 
$\theta$=0$^o$, 45$^o$ and 150$^o$. Note that the ratios are almost the same 
for small emission angles $0{\le}\theta{\le}45^o$.  But the ratio becomes large
at  $\theta$=150$^o$. 

In addition Fig.~\ref{neff5c}b) illustrates the uncertainty in the evaluation of
$\sigma_f$ from the ratios of data for $f$ particle production from different
nuclear targets. Namely, it is clear that for different emission angles the
$N_{Pb}{/}N_C$ ratios roughly saturate at $\sigma_f{\ge}$40~mb. This actually
means that model is insensitive to the  value of $\sigma_f$ if it exceeds the
limit of $\simeq$40~mb.

For completeness, Fig.~\ref{cofi10} illustrates the slope $\alpha$ of the
$A$-dependence given by Eq.~(\ref{slope}). The results are shown for the
different production angles and as a function of $\sigma_f$. The calculations
were done with $\sigma_i$=40 mb. Note that $\alpha$ is saturated above
$\sigma_f{\simeq}$40 mb, while the slope varies significantly at $\sigma_f{\le}$
20~mb. The variation of $\alpha$ with the emission angle is almost negligible at
$\theta{\le}45^0$. Because of the distortion of the incident particle the
maximal value of $\alpha$ is below one. The minimal slope is close to
$\simeq$0.3.

\section{{\boldmath$A$}-dependence due to two-step production}

Fig.~\ref{cofi10} shows that, neglecting the distortion of the final particle,
{\it i. e.} for $\sigma_f{=}0$,  one might expect the maximal value for  the
slope of the $A^\alpha$-dependence around $\simeq$0.7, which is essentially
driven by the distortion of the incident particle given by  $\sigma_i$ used in
our calculations. Considering proton-nucleus interactions one can use
$\sigma_i$=40 mb in the range of proton energies from $\simeq$3 to 10$^3$ GeV. 
However, many experiments~\cite{Abt,Antipov1,Antipov2,Dijkstra,Daum,Bailey} done
with high energy proton beams indicate a slope $\alpha{\simeq}1$. This is the
case for $\phi$-meson production in $pA$ collisions.

This apparent discrepancy can be understood quantitatively in terms of
multi-particle production at high energies. The possible scenario is production
of many pions that interact inside the nucleus and produce $\phi$-mesons. Since
the flux density of these pions or their multiplicity can be large and their
energies are above the $\pi{N}{\to}\phi{N}$ threshold, the probability of this
process could be larger than the probability of the direct production considered
previously. Indeed the pions are distributed through the whole nucleus and thus
the $\phi$-meson can be produced over the full volume of the target. Therefore
one can expect that the $A$-dependence of the $\phi$-meson production in such a
case is proportional to $A$, while neglecting the final distortion. 

Quantitative estimates can be done for the two-step process. Assume that at some
impact parameter  ${\bf b}$ and longitudinal point $\tilde{z}$ the incident
energetic particle $i$ produces some intermediate state $j$. Due to the Lorentz
boost, the $j$ particle is moving along the beam direction, {\it i. e.} at the
same impact parameter ${\bf b}$ and at some point $z$ produces the final
particle $f$. The final particle is now moving along the path given by the polar
angle $\theta$ and azimuthal angle $\phi$ of the emission. 

Thus we have two sub-processes and this is called a two-step mechanism. The
first one is $iN{\to}jN$ and the second one is $jN{\to}fN$ where the distortion
of the $i$, $j$ and $f$ particles is taken into account. The probability of the
first process can be derived in analogy to Eq.~(\ref{class1}) and is given as
\begin{eqnarray}
S_{ij}({\bf b},z)= |f_{ij}|^2 \!\!\int\limits_{-\infty}^z \!\!d{\tilde z}\ 
\rho_A({\bf b},{\tilde z}) \nonumber \\ \exp\left[{-}
\sigma_i\!\!\!\int\limits_{-\infty}^{\tilde z} \!\!dz'\,
\rho_A({\bf b},z')
- \sigma_j\!\!\!\int\limits_{\tilde z}^{z} \!\!dz'\,
\rho_A({\bf b},z')
\right]\!\!,
\label{two1}
\end{eqnarray}
where $f_{ij}$ is the amplitude of the $iN{\to}jN$ transition, while $\sigma_i$
and $\sigma_j$ account for the distortion of the $i$ and $j$ particle. The
attenuation probability of passing $f$ through the nucleus is similar to
Eq.~(\ref{sf}). 

The effective collision number for the two-step process is 
\begin{eqnarray}
N_A=\!\!\int\! d^2b \, dz\, \rho_A({\bf b},z)
\,S_{ij}({\bf b},z) \,S_f({\bf b},z,\theta,\phi ),
\label{twotwo}
\end{eqnarray}
where $S_f$ is given by Eq.~(\ref{sf}). Note that for $S_{ij}{=}S_f{=}1$ the
$A$-dependence is proportional to $A$.

\begin{figure}[t]
\vspace*{-3mm}
\centerline{\hspace*{1mm}\psfig{file=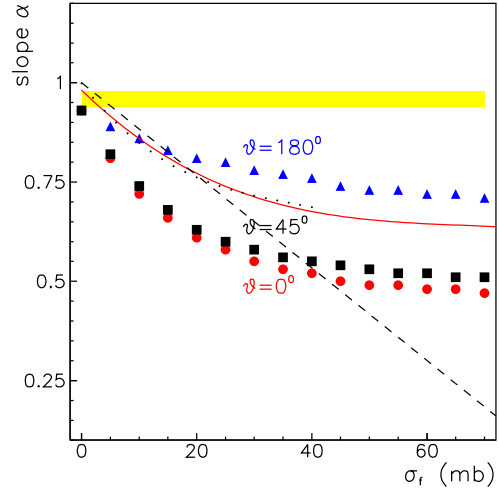,width=7.5cm}}
\vspace*{-3mm}
\caption{The slope of the $A^\alpha$-dependence as a function of the effective
cross section $\sigma_f$ shown for the different production angles:
$\theta{=}0^0$ (circles), $\theta{=}45^0$ (squares) and $\theta{=}180^0$
(triangles). The results are shown for the two step production mechanism and the
calculations were done using Eq.~(\ref{twotwo}). The solid line indicate the
results obtained by Eq.~(\ref{hi}). The dashed line is the  $\langle\rho
L\rangle$ approximation calculated for $Ag$ nucleus. The shaded box indicates
the experimental result from the HERA-B Collaboration.}
\label{cofi10a}
\end{figure}

Now Fig.~\ref{cofi10a} shows the slope $\alpha$ of the $A^\alpha$-dependence as
a function of the effective cross section $\sigma_f$. The calculations were done
using Eq.~(\ref{twotwo}) and the results are shown for the different production
angles. Note that the slope for the two-step production mechanism can be
substantially larger than for the direct one.  For the two-step process $\alpha$
can be close to one. However one can not estimate the $A$-dependence due to the
multi-step production and thus the extraction of $\sigma_f$ can not be done
unambiguously.

A quite  different  estimate of the $A$-dependence was proposed in
Ref.~\cite{Kharzeev} for the evaluation of charmonium absorption from high
energy $J/\Psi$-production in $p{+}A$ collisions. In that case the
$A$-dependence is given by
\begin{eqnarray}
N_A{=}\!\!\int\!\!\! d^2b \, dz\, \rho_A({\bf b},z)
\exp\!\!\left[{-}\frac{A{-}1}{A}\,
\sigma_f\!\!\!\int\limits_{z}^{\infty} \!\!dz'\,
\rho_A({\bf b},z')\right].
\label{hi}
\end{eqnarray}
Following this formalism it is clear that if there is no distortion of the final
particle the $A$-dependence is proportional to $A$ and this is the limit, which
might  be expected from the nuclear data on incoherent particle production. The
solid line in Fig.~\ref{cofi10a} indicates the slope $\alpha$ of the $A^\alpha$
dependence calculated by Eq.~(\ref{hi}) for different $\sigma_f$. Note that in
the derivation of Eq.~(\ref{hi}) it was assumed~\cite{Kharzeev} that $\sigma_f$
is small, so one can not seriously discuss the difference between the two-step
formalism and the above estimates at large distortion cross sections.

Another simple evaluation of the $A$-dependence can be done using so-called
$\rho{L}$ parameterization~\cite{Alessandro}
 \begin{eqnarray} N_A=A \exp
\left[{-\sigma_f \, \langle \rho L \rangle} \right],
\end{eqnarray}
where $\langle\rho L\rangle$ is the average amount of matter crossed by the
final particle and\footnote{Note that our normalization of the nuclear density
is $A$, while in Refs.~\cite{Kharzeev,Alessandro} it is unity.}
\begin{eqnarray}
\langle\rho L \rangle= \frac{A-1}{2A^2}\int d^2b [T({\bf b})]^2.
\end{eqnarray}

Now if $\sigma_f$ is small one can use the following expression
\begin{eqnarray}
N_A=A^\alpha, \,\,\, \alpha = 1-\sigma_f\frac{\langle\rho L \rangle}{\ln A},
\label{rl}
\end{eqnarray}
which was extensively  applied in the evaluation of the distortion  of the 
charmonium cross section in nuclear matter. 

Apparently the slope $\alpha$ can be calculated using only one target and as we
found only slightly depends on $A$ unless one uses light targets. The dashed
line in Fig.~\ref{cofi10a} shows the $\alpha$ obtained by this $\langle\rho
L\rangle$ approximation using Eq.~(\ref{rl}) and an $Ag$ target. As was
mentioned before, the approximation is valid for small $\sigma_f$ and indeed is
in rough agreement with Eq.~(\ref{hi}) for $\sigma_f{\le}$20 mb.

To obtain experimental values for the slope $\alpha$ one needs the production
cross sections measured for different nuclear targets $A$. It was
found~\cite{Alessandro} in the evaluation of charmonium absorption, that the
value of $\alpha$ extracted from a fit to a given data set depends on the
nucleus used as the lightest target. Indeed the experiments that use heavy
targets with hydrogen or deuterium systematically obtain large values of the
slope $\alpha$.

\section{{\boldmath$\phi$}-meson production in {\boldmath$p{+}A$} collisions at
high energies}

As was mentioned  previously the results on $A$-dependence of inclusive
$\phi$-meson production from $p{+}A$ collisions at high beam energies indicate a
large slope $\alpha{\simeq}1$. Here we shortly review the current status.
Moreover, we evaluate the $\phi$-meson distortion cross section and collect the
results in Table~\ref{tab:data-1}. Furthermore the interpretation of
$\sigma_\phi$ is given in the next Section.

Most recently the $A$-dependence of inclusive $\phi$-meson production off nuclei
using  a 920 GeV proton beam was measured with HERA-B detector at HERA storage
ring~\cite{Abt}. This experiment was done with $C$, $Ti$ and $W$ targets and the
$\phi{\to}K^+K^-$ decay mode was used for the $\phi$-meson reconstruction. The
data analysis shows the slope $\alpha{=}0.96{\pm}0.02$. As discussed before only
the multi-step mechanism can be an explanation of the HERA-B observation. Indeed
the shaded box in Fig.\ref{cofi10a} indicates the result from the HERA-B
Collaboration, which in principle can be explained assuming
$\sigma_\phi{=}2.1{\pm}1.2$ mb.  Here we use Eq.~(\ref{hi}) as was done in the
analysis of $J/\Psi$ distortion. 

A systematic study of $\phi$-meson production in $p{+}A$ collisions at a beam
energy of 12 GeV was carried ot  by the KEK-PS E325
Collaboration~\cite{Ozawa,Tabaru,Muto,Sakuma}. The slope $\alpha$ of the
$A^\alpha$-dependence was evaluated using $C$ and $Cu$ targets. The most recent
results~\cite{Sakuma} allows to investigate how $\alpha$ depends on the
$\phi$-meson momentum as well as to obtain $\alpha$ for the $\phi{\to}K^+K^-$
and $\phi{\to}e^+e^-$ decay mode. It was found that $\alpha$ is statistically
the same for these two different decay modes in the same kinematical region.
Furthermore it turns out that $\alpha$ depends on reaction kinematics. Here we
would like to make some comments.

The relevant kinematics for the evaluation of A-dependence is given by the final
particle production angle. It is clear that this angle defines the path of the
particle and therefore the amount of matter involved in the distortion.
Eq.~(\ref{hi}) is applicable at forward angles, while Eq.~(\ref{twotwo}) can be
used for large angles but accounts only for the two-step production mechanism.
Nevertheless within such limitations one can realize from Fig.~\ref{cofi10a}
that the angular dependence is essential for data evaluation. 

Unfortunately, the KEK-PS E325 data are given either as a function of
$\phi$-meson momentum alone~\cite{Sakuma} or as a function of rapidity and
transverse momentum~\cite{Tabaru}. In our opinion, the ideal case is to fix
forward angles and to extract the slope $\alpha$ for the different laboratory
momenta of the produced $\phi$-mesons. That would give information about the
momentum dependence of the distortion. In spite of that uncertainty in the
analysis of the KEK-PS E325 data our results for the $\phi$-meson distortion
cross section are summarized in the Table~\ref{tab:data-1}.

The $A$-dependence of the inclusive $\phi$-meson production in neutron-nucleus
interactions at 30-70 GeV was studied by the BIS-2 Collaboration at the
Serpukhov accelerator~\cite{Aleev}. Here the $C$, $Al$ and $Cu$ targets were
used and it was found that the slope $\alpha=0.81{\pm}0.06$. That corresponds to
a distortion cross section of $12{\pm}4$ mb.

In Ref.~\cite{Daum}, the $A$-dependence of $\phi$ meson production by a 100
GeV/c proton beam was determined through the analysis of the data collected with
$H_2$ and $Be$ targets. The measurements was done by the ACCMOR Collaboration at
SPS. The $\phi{\to}K^+K^-$ decay mode was used for the reconstruction. It was
found that slope $\alpha{=}0.96{\pm}0.04$. Moreover, it was argued~\cite{Daum}
that the use of the $H_2$ target in general introduces additional systematic
uncertainties, which are difficult to estimate. This result is close to the
HERA-B observation.

\begin{table}[t]
\begin{center}
\caption{The slope $\alpha$ of the $A^\alpha$-dependence of $\phi$-meson
production obtained in the different experiments and the distortion cross
section $\sigma_\phi$ evaluated by Eq.~(\ref{hi}). For the KEK-PS E325 results
the first error in $\alpha$ is statistical and the second error is systematic.
The results are shown for the different ranges of rapidity $y$ and transverse
momentum $p_t$ given in GeV/c.}
\label{tab:data-1}
\begin{tabular}{|c|r|r|c|c|}
\hline\noalign{\smallskip}
\multicolumn{2}{|c|}{Experiment}& Ref. & $\alpha$ & $\sigma_\phi$ (mb) \\
\noalign{\smallskip}\hline\noalign{\smallskip}
\multicolumn{2}{|c|}{ HERA-B} & \cite{Abt} & 0.96$\pm$0.02 & 2.1$\pm$1.2  \\
\multicolumn{2}{|c|}{BIS-2} & \cite{Aleev} & 0.81$\pm$0.06 & 12$\pm$4 \\
\multicolumn{2}{|c|}{ ACCMOR} &\cite{Daum} & 0.96$\pm$0.04 & 2.1$\pm$2 \\
\multicolumn{2}{|c|}{NA 11} & \cite{Bailey} & 0.86$\pm$0.02 & 9$\pm$2 \\
\noalign{\smallskip}\hline\noalign{\smallskip}
\multicolumn{2}{|c|}{KEK-PS E325} & \cite{Tabaru} & & \\
$y$ & 0.9-1.1 &  & 0.916$\pm$0.101$\pm$0.022 \,\,\,&4.9$\pm$4 \\
$y$ & 1.1-1.3 &  & 1.050$\pm$0.101$\pm$0.02 & 0$\pm$2.8 \\
$y$ & 1.3-1.5 & & 0.881$\pm$0.084$\pm$0.02 & 7.2$\pm$5.8 \\
$y$ & 1.5-1.7 & & 0.780$\pm$0.119$\pm$0.019 & 14$\pm$8.3 \\
$p_t$  & 0-0.25&  & 0.971$\pm$0.101$\pm$0.019 & 1.7$\pm$7  \\
$p_t$ & 0.25-0.50 & & 0.890$\pm$0.066$\pm$0.019 & 6.7$\pm$4.9 \\
$p_t$ & 0.50-0.75 & & 0.924$\pm$0.111$\pm$0.021& 4.4$\pm$4  \\
\noalign{\smallskip}\hline\noalign{\smallskip}
\end{tabular}
\end{center}
\end{table}

The $A$-dependence of the inclusive $\phi$-meson production from beryllium and
tantalum targets using  a 120 GeV proton beam was studied with the NA11
spectrometer at CERN SPS~\cite{Bailey}. The data analysis indicates that
$\alpha{=}0.86{\pm}0.02$.  Applying Eq.~(\ref{hi}) one can estimate
$\sigma_\phi{\simeq}9{\pm}2$ mb. 

Below we also list experiments that did not measure the $A$-dependence, but
assume some values of $\alpha$ under certain assumptions in order to analyze the
data.

Inclusive $\phi$-meson production off beryllium nuclei by 70 GeV/c protons was
studied with the Sigma spectrometer~\cite{Antipov1} at the Serpukhov
accelerator. For the evaluation of elementary $pN{\to}\phi{X}$ cross section it
was assumed that the $A$-dependence of the nuclear cross section is proportional
to $A^{0.7}$, as was measured for $K^-A$ interactions~\cite{Antipov2}.  

High statistics $\phi$-meson production from $p{+}Be$ collisions at beam momenta
of 120 and 200 GeV/c was studied by the ACCMOR Collaboration at
SPS~\cite{Dijkstra}. The data evaluation was done under the assumption that the
$A$-dependence is proportional to $A$, which was motivated by the experimental
results published in Refs.~\cite{Daum,Bailey}. 

Finally the evaluated distortion cross sections are collected in
Table~\ref{tab:data-1}. Unfortunately large uncertainties in the experimental
results for $\alpha$ produce large uncertainties in $\sigma_\phi$. In our
opinion, the analysis of the ratios of $\phi$ production cross sections from
different nuclei with respect to the $C$-target results might be less uncertain.
In that case the systematical errors might cancel up to large extent. However,
such an analysis requires measurements with many different nuclear targets,
which is not the case for some experiments available now.

\section{Interpretation of {\boldmath$\sigma_f$}}

\subsection{Definitions}

The interpretation of $\sigma_i$ is a general problem. Following our derivation
given in Section~\ref{formal}, the eikonal formalism operates with \, the
forward scattering \, amplitude $f(0)$ and \, $\sigma_f$ appears through the
optical theorem. In that sense $\sigma_f$ is the total cross section for the
interaction of a particle $f$ with a nucleon embedded in nuclear matter. In the
classical derivation $\sigma_f$ is considered a distortion cross section. There
is no conflict between these two definitions if we consider attenuation of the
flux of the final particle due to all possible processes available in the
nucleus. That might be  general absorption, scattering out of the initial
trajectory, decay of an unstable particle followed by the distortion of the
decay products, interaction with few nucleon configurations and whatever one can
assume. The total sum over all these possible processes is an effective total or
distortion cross section. 

As we emphasized previously  $\sigma_f$ is not the free vacuum $f{+}N$ total
cross section since it has to be extracted from the nuclear data and might be
modified by in-medium effects. However it is always worthwhile to compare
$\sigma_f$  with the free cross section if that is available. That comparison
would show whether additional reaction channels were open in nuclear matter or
whether reaction channels available in free space are blocked in the nucleus.
For instance some transitions might be blocked  due to the Pauli principle. In
the analysis of charmonium properties in nuclear matter, the distortion cross
section $\sigma_f$ is a standard variable generally used everywhere throughout
the relevant discussions.

Since in some calculations not the distortion cross section but other variables
are used we provide here  some useful relations for the conversion. Let us first
remind the reader that the complex forward scattering amplitude is related to
the cross section by Eq.~(\ref{forward}). At the same time the complex local 
potential is given in terms of the complex forward scattering amplitude $f(0)$
as~\cite{Dover,Gopal,Rosental,Voloshin}
\begin{eqnarray}
V=- 2\pi \, \frac{m_N{+}m_f}{m_Nm_f} \, \rho \, f(0),
\end{eqnarray}
where $\rho$ is local nuclear density.  The potential depends on $f$ due to the
energy dependence of $f(0)$. It is possible to use a so called in-medium
collisional width $\Delta\Gamma$ and mass shift $\Delta{m}$ of the $f$ particle,
which are~\cite{Lenz,Dover1,Friman,Klingl}
\begin{eqnarray}
\Delta\Gamma = 4\pi \frac{m_N{+}m_f}{m_Nm_f} \rho \, {\rm Im} f(0)
{=}\frac{m_N{+}m_f}{m_Nm_f} \rho \, k_f  \,  \sigma_f  \label{width} \\
\Delta{m} = -2\pi \frac{m_N+m_f}{m_Nm_f} \rho \, {\rm Re} f(0),
\label{mass}
\end{eqnarray}
where $m_N$ is the nucleon mass and $k_f$ is the momentum of the final
particle.\footnote{It is clear that the introduction of $\Delta\Gamma$ in the
calculation described above requires an accurate definition of the nuclear
density dependence rather then average estimate of Eq.~(\ref{width}).} In
principle one can replace masses by total energies and discuss the
$\Delta\Gamma$ and $\Delta{m}$ at high energy of the final particle.  Note that
the in-medium collisional width and mass shift are not invariants and can be
changed by a Lorentz boost, so one should use these variables in the rest frame
of the $f$ particle.

\subsection{Estimates for {\boldmath$\sigma_\phi$} in vacuum}

It is useful to compare distortion cross sections extracted from the nuclear
data with its values in free space. That allows to inspect directly  the
possible in-medium modification of the  $\sigma_\phi$. There are various well
known methods to estimate the $\phi{+}N$ interaction cross section.

The $\phi{+}N$ cross section can be evaluated in the Vector Dominance Model from
the $\gamma{N}{\to}\phi{N}$ reaction. Within VDM the hadron-like
photon~\cite{Stodolsky} is a  superposition of all possible vector meson states.
Therefore the  $\gamma{N}{\to}\phi{N}$ reaction can be decomposed into the
transition of the photon to a virtual vector meson $V$ followed by the elastic
or inelastic vector meson scattering on the target nucleon and production of the
final $\phi$-meson.  The reaction amplitude is then written as~\cite{Bauer,Paul}
\begin{eqnarray}
f_{\gamma{N}{\to}\phi{N}} = \sum_{V}\frac{\sqrt{\pi\alpha}}{\gamma_V}
f_{VN{\to}\phi N},
\label{ampli1}
\end{eqnarray}
where the summation is performed over vector meson states. Moreover, $\alpha$ is
the fine structure constant, $\gamma_V$ is the photon coupling to the vector
meson $V$ and $f_{VN{\to}\phi N}$ is the amplitude for the $VN{\to}\phi{N}$
transition.

The coupling $\gamma_V$ is given by vector meson decay into a lepton
pair~\cite{Nambu}
\begin{eqnarray}
\Gamma(V{\to}l^+l^-)=\frac{\pi\alpha^2}{3\gamma_V^2}\sqrt{m_V^2-4m_l^2}
\left[1+\frac{2m_l^2}{m_V^2}\right],
\label{coupl1}
\end{eqnarray}
where $m_V$ and $m_l$ are the masses of vector meson and lepton, respectively.
Taking the di-electron decay widths~\cite{PDG}, the photon couplings to the
lightest vector mesons are
\begin{eqnarray}
\gamma_\rho{=}2.51, \,\,\,\,\,\,
\gamma_\omega{=}8.47, \,\,\,\,\,\,
\gamma_\phi{=}6.69.
\label{coupl2}
\end{eqnarray}
Note that non-diagonal, {\it i.e.} $\rho{N}{\to}\phi{N}$ and
$\omega{N}{\to}\phi{N}$ as well as  diagonal $\phi{N}{\to}\phi{N}$ transitions
contribute to the reaction amplitude of Eq.~(\ref{ampli1}). VDM suggests that
the virtual vector meson stemming from the photon becomes real through the
four-momentum $t$ transferred to the nucleon, which in general requires the
introduction of a form-factor at the interaction
vertices~\cite{Hufner,Sibirtsev11,Sibirtsev12}. In many analyses~\cite{Bauer},
this form factor is neglected. Thus the VDM analysis of photoproduction data
requires additional assumptions. 

\begin{figure}[t]
\vspace*{-3mm}
\centerline{\hspace*{3mm}\psfig{file=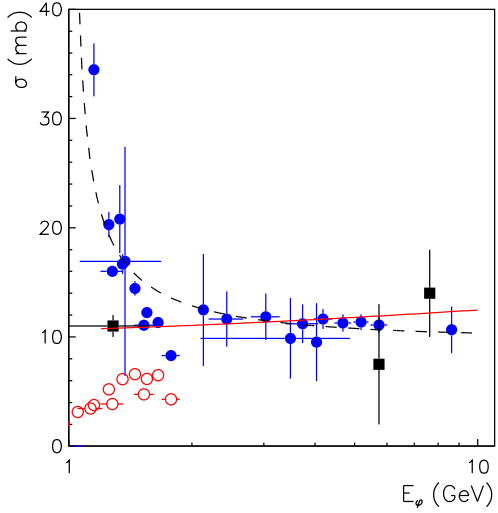,width=7.5cm}}
\vspace*{-2mm}
\caption{The $\phi{N}$ cross section as a function of $\phi$-meson total energy.
The open circles show the results obtained by Eqs.~(\ref{vdm1a}, \ref{tocro})
from the data on forward $\phi$-meson photoproduction cross section assuming
$s$-wave dominance in the $\phi{N}{\to}\phi{N}$ scattering. The closed circles
are the estimates for $\sigma_{\phi{N}}{\times}\sqrt{1+\alpha_\phi^2}$ given by
Eq.~(\ref{vdm1}). The squares are the results extracted~\cite{Sibirtsev1} from
the data on $\phi$-photoproduction from nuclei~\cite{Mcclellan,Ishikawa}. The
solid line is constrained by the additive quark model of Eq.~(\ref{aqm}). The
dashed line shows the estimate using Eq.~(\ref{fittoc}).}
\label{cofi11}
\end{figure}

There are many precise data on $\phi$-meson photoproduction differential cross
sections at energies close to the reaction threshold. These data can be used for
evaluation of the $\phi{N}{\to}\phi{N}$ scattering amplitude squared applying
the VDM as
\begin{eqnarray}  
\frac{d\sigma_{\gamma N{\to}\phi N}} {dt}
=\frac{\pi^2\, \alpha}{\gamma_\phi^2 \, q^2_\gamma}
\left|f_{\phi{N}{\to}\phi{N}}\right|^2.
\label{vdm1a}
\end{eqnarray}
Now if the scattering is dominated by $s$-waves one could extract the $\phi{N}$
scattering length at threshold. It is related to the cross section
as\footnote{In our normalization the scattering amplitude $f$ equals to the
scattering length $a_{\phi{N}}$ at  $q_\phi{\to}0$.}
\begin{eqnarray}  
\sigma_{\phi N}=4\pi \left| f_{\phi{N}{\to}\phi{N}} \right|^2
\label{tocro}
\end{eqnarray}
and is shown in Fig.~\ref{cofi11} by open circles. Here we use the forward
$\phi$-meson photoproduction cross section available near the reaction
threshold~\cite{Barth,Mibe}. This scattering length can be compared with other
theoretical predictions. For instance the estimate based on QCD sum
rules~\cite{Koike} provides  a real $\phi{N}$ scattering length of
$a_{\phi{N}}{\simeq}$-0.15 fm that corresponds to a cross section of
$\sigma_\phi{\simeq}$2.8 mb and seems to be in good agreement with the data
evaluated by Eqs.~(\ref{vdm1a}, \ref{tocro}) under assumption of $s$-wave
$\phi{N}$ scattering. However, note that the $\phi$-meson photoproduction
differential cross sections are essentially anisotropic already at energies
close to the threshold and can be well parametrized as
$d\sigma{/}dt{\propto}\exp{(bt)}$ with a slope $b{\simeq}$3 GeV$^{-2}$ and $t$
being the four momentum transfer squared~\cite{Barth,Mibe}. In that sense,  the
estimates shown by an open circles in Fig.~\ref{cofi11} might not be  correct
and should  be taken with a grain of salt in  the evaluation of  $\phi{N}$
scattering length.

The estimate~\cite{Klingl1} based on an effective Lagrangian approach predicts
the scattering length $a_{\phi{N}}{=}(-0.01{+}i0.08)$~fm that corresponds to
$\sigma_\phi{\simeq}$0.8 mb.

Another estimate can be obtained from a  QCD van der Waals potential
calculation~\cite{Gao},  which predicts a $\phi$-nucleon bound state. In that
case the Born approximate scattering length is given by the potential
as~\cite{Brodsky,Brodsky1}
\begin{eqnarray}
a_{\phi{N}} =2 \frac{m_N \, m_\phi}{m_N + m_\phi} \int\limits_0^\infty dr \, 
r^2 \, V(r),
\end{eqnarray}
where the potential was taken in the Yukawa form $V(r)={-}\alpha\exp[-r\mu]{/}r$
with strength $\alpha{=}1.25$ and  range $\mu{=}$0.6 GeV. These parameters were
obtained for an attractive potential and result in $a_{\phi{N}}{\simeq}$0.67 fm.
They correspond to $\sigma_{\phi}$=56 mb at $q_\phi$=0.

Furthermore, applying Eq.~(\ref{optic1})  the $\gamma{N}{\to}\phi{N}$
differential cross section of Eq~.(\ref{vdm1a}) at $t{=}0$ can be written as
\begin{eqnarray}  
\left.{\frac{d\sigma_{\gamma N{\to}\phi N}}{dt}}\right|_{t{=}0}\!\!\!\!\!
=\frac{\alpha}{16\gamma_\phi^2} \, \frac{q^2_\phi}{q^2_\gamma}\,
(1+\alpha_\phi^2)\, \sigma_{\phi N}^2.
\label{vdm1}
\end{eqnarray}

Since the ratio of the $\alpha_\phi$ and $\phi{N}$ total cross sections are
unknown, one can extract from the photoproduction data only their combination,
{\it i. e.} $\sigma_{\phi{N}}{\times}\sqrt{1+\alpha_\phi^2}$, which is shown by
closed circles in the Fig.\ref{cofi11}. Here we use the data collected in
Ref.~\cite{Sibirtsev1}. If one assumes that $\alpha_\phi{=}0$ these results may
be considered as the energy dependence of the $\phi{N}$ cross section. While for
many processes the real part of the scattering amplitude vanishes at high
energies, the ratio of real to imaginary part of the amplitude $\alpha$ is large
at low energies and moreover substantially depends on the momentum of the
scattered particle~\cite{PDG}. Just to illustrate such a possibility the dashed
line in Fig.~\ref {cofi11} shows the dependence
\begin{eqnarray}
\sigma_{\phi N}{\times}\sqrt{1+\alpha_\phi^2} =
10~{\rm (mb)} \times \sqrt{1+\frac{0.6~{\rm (GeV/c)} }{q_\phi^2}}.
\label{fittoc}
\end{eqnarray}
Again this might ensure that at high energies the $\phi{N}$ cross section
approaches some value around 10~mb, but still does not provide a trustworthy
estimate of $\sigma_{\phi{N}}$ close to threshold. At least it is not
appropriate to estimate the real part of the forward scattering amplitude and to
evaluate the in-medium mass shift using Eq.~(\ref{mass}).

Within an additive quark model the $\phi{N}$ cross section is given
as~\cite{Lipkin}
\begin{eqnarray}
\sigma_{\phi N} =\sigma_{K^-N} +\sigma_{K^+N} - \sigma_{\pi^-N},
\label{aqm}
\end{eqnarray}
where the  elementary cross sections are taken at the same invariant collision
energies. Since the $\phi{N}$ reaction threshold is $m_\phi{+}m_N{\simeq}$1.96
GeV one can safely use Regge parametrization for the continuum or non-resonant
meson-nucleon scattering amplitudes~\cite{Cudell}. Now Eq.~(\ref{aqm}) is shown
by the solid line in Fig.~\ref{cofi11} and is in reasonable agreement with VDM
results at high energies.

It is also worthwhile to show the estimate based on  the dynamical study of the
$\phi{N}$ bound state within the chiral SU(3) quark model.  By solving a
resonating group method based equation \cite{Huang} it was found that the
binding energy of the state might range from 1 to 9 MeV. With respect to the
$s$-wave scattering length $a_{\phi{N}}$, the relation between the pole of the
$S$-matrix and binding energy $\epsilon$ is given as
\begin{eqnarray}
a_{\phi N}=\left[\frac{2 m_N \, m_\phi}{m_N +m_\phi}\,  \epsilon \right]^{-1/2},
\end{eqnarray}
so that the real part of the scattering length ranges from 2.1 to 6.3~fm. This
scattering length is large compared to the other results. Note that this result
is used in three-body  calculations of the $\phi{NN}$ nuclear cluster binding
energy~\cite{Belyaev}.
	
\subsection{Estimates for {\boldmath$\sigma_\phi$} in matter}

Only some results evaluated  from high energy proton-nucleus collisions
(summarized in Table~4) are in agreement with the data shown in
Fig.~\ref{cofi11}. The uncertainties of the KEK-PS data~\cite{Tabaru} are still
too large to draw a definite conclusion right now. The results from
BIS-2~\cite{Aleev} and NA-11~\cite{Bailey} are consistent with vacuum estimates
at high energies.

However the results from HERA-B~\cite{Abt} and ACCMOR~\cite{Daum} indicate a
substantially smaller $\phi$-meson distortion cross section. This observation is
difficult to interpret since in high energy experiments, the $\phi$-mesons are
produced with high momenta and should be almost blind to any in-medium
modification. 

The squares in Fig.~\ref{cofi11} show the $\phi$-meson distortion cross section
extracted in~\cite{Sibirtsev1}  from the data on $\phi$-photoproduction from
nuclei~\cite{Mcclellan,Ishikawa}. As we already discussed, it is not necessary
that these in-medium results are the same as $\sigma_{\phi{N}}$ in vacuum.
However we observe reasonable agreement between nuclear results and those
evaluated by VDM at high energies. 

Substantial modification of slow $\phi$-mesons in nuclear matter was proposed in
Ref.\cite{Klingl1}. It was found that the mass of the $\phi$-meson almost does
not change in matter, while the change  of the width accounts for
$\Delta\Gamma{\simeq}$45~MeV at normal nuclear density. Following
Eq.~(\ref{width}) one can estimate the distortion cross section as
$\sigma_\phi{\simeq}$70 mb for $k_\phi$=100 MeV/c.

The energy dependence of the in-medium $\phi$-meson width was studied in
Ref.~\cite{Oset1}. While for $E_\phi{=}m_\phi$  the width is about 20 MeV at
normal nuclear density, it increases up to 40 MeV at a $\phi$-meson energy of
1.1 GeV. So it is really changed by a factor of two over 80 MeV in
energy.\footnote{Here we refer to the energy dependence of the $\phi$-meson
width in nuclear matter at different densities, which is shown in the Fig.4 of
Ref.~\cite{Oset1}.} This corresponds to a variation of the distortion cross
section from $\simeq$27 to 15 mb and seems to be in agreement with the dashed
line shown in the Fig.~\ref{cofi11}.

Furthermore, the  energy dependence of the $\phi$-meson width at normal nuclear
density was investigated in Ref.~\cite{Cabrera}. It was found that the in-medium
width $\Delta\Gamma{\simeq}$22-17~MeV slightly varies with energy within the
range $E_\phi{=}m_\phi$ to 1.2 GeV.

\section{Predictions for {\boldmath$\phi$}-meson production in
{\boldmath$p{+}A$} collisions at COSY energies}

The COoler SYnchrotron (COSY) at J{\"u}lich provides an unique opportunity to
study the $\phi$-meson distortion in nuclear matter at low energies. Our
analysis indicates that even the vacuum $\phi{N}$ interaction is not well
understood  and different estimates illustrated in Fig.~\ref{cofi11} are in
substantial disagreement at $\phi$-meson energies below  2 GeV. Moreover, the
available predictions~\cite{Klingl1,Oset1,Cabrera} state that the in-medium
modification of  the $\phi$-meson width is substantial at low energies, although
the real size of that change is not well established. 

Such a situation requires precise measurements of $\phi$-meson production from
$p{+}A$ collisions at low energies, as was proposed in
Refs.~\cite{Magas1,Magas2}. A dedicated experiment on $\phi$-meson production
from  the proton interaction with $^{12}C$, $^{108}Ag$ and $^{197}Au$ targets at
maximum COSY energies of 2.83 GeV was proposed by ANKE
Collaboration~\cite{Hartmann}. Here we show the results for the $A$-dependence
of $\phi$-meson production in $pA$ collisions at few GeV energies. Note that at
low energies the $\phi$-meson production due to multiple processes is suppressed
due to the final particle multiplicities and large $\phi$-meson production
threshold.

\begin{figure}[t]
\vspace*{-3mm}
\centerline{\hspace*{3mm}\psfig{file=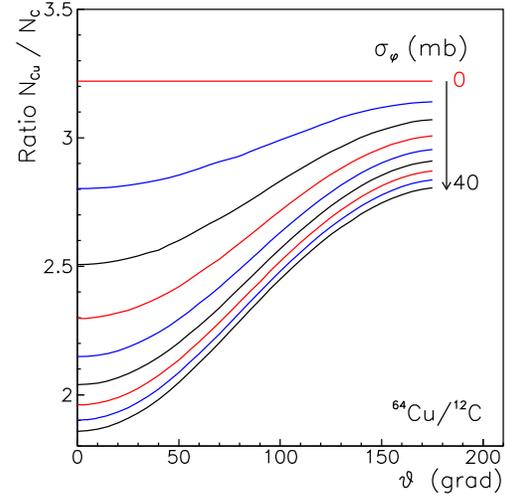,width=7.5cm}}
\vspace*{-3mm}
\caption{ The ratio of the effective collision numbers calculated for $^{64}Cu$
and $^{12}C$ nuclear targets as a function of $\phi$-meson production angle
shown for the different $\sigma_\phi$ within the range from 0 to 40 mb with a
step size of 5 mb.}
\label{cofi4a}
\end{figure}

For the further calculations we fix $\sigma_i$=40~mb, which stands for the
average cross section for the interaction of the beam protons with target proton
and neutron. Although the $pN$ interaction can be modified in nuclear matter one
would not expect that this effect is significant for the protons with momenta
above $\simeq$1~GeV/c.

We believe that the analysis of the ratios $R$ of the produced $\phi$-meson
contains less theoretical uncertainties than the analysis of the differential
cross section $d^2\sigma{/}dT{/}d\Omega$ itself for each nuclear target $A$ or
the slope $\alpha$, as was discussed previously. Thus in the following we show
our predictions for the ratio
\begin{eqnarray}
R =\frac{d^2\sigma_A}{dT\, d\cos\theta}\times\left[\frac{d^2\sigma_C}{dT\,
d\cos\theta}\right]^{-1}
=\frac{N_A}{N_C}
\end{eqnarray}
taken with respect to the carbon target. Here $T$ is the kinetic energy and
$\theta$ is the emission angle of the produced $\phi$-meson, while $N_A$ is the
effective collision number that was calculated for the different $\sigma_\phi$
and $\theta$.  Moreover, the analysis of the ratios has additional advantages
since systematic experimental uncertainties can be substantially reduced.

\begin{figure}[b]
\vspace*{-3mm}
\centerline{\hspace*{3mm}\psfig{file=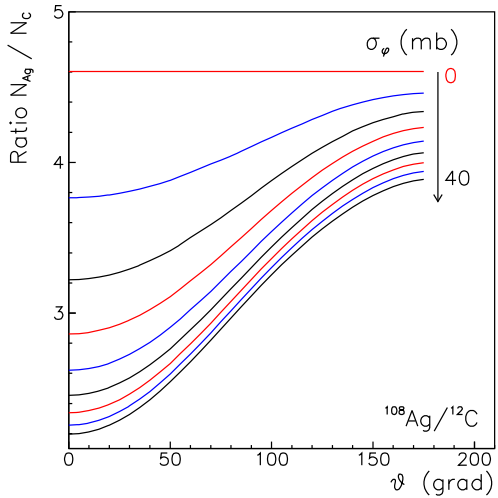,width=7.5cm}}
\vspace*{-3mm}
\caption{The ratio of the effective collision numbers calculated for $^{108}Ag$
and $^{12}C$ targets as a function of the $\phi$-meson production angle shown
for different $\sigma_\phi$ within the range from 0 to 40 mb with a step size of
5 mb.}
\label{cofi4}
\end{figure}

Once more we would like to emphasize that in the evaluation of an effective
collision number $N_A$ we use an effective in-medium cross section
$\sigma_f{=}\sigma_\phi$. This is not the cross section for the $\phi$-meson
interaction with free nucleon. Moreover, as was discussed previously the
multiple scattering series corrections $\epsilon$ given by Eq.~(\ref{xsect1})
can not be isolated and thus the extracted $\sigma_\phi$ cross section contains
such a multiple scattering contribution. Nevertheless it is of great importance
to compare $\sigma_\phi$ evaluated from the nuclear data with the vacuum
$\phi{N}$ cross section.

\begin{figure}[t]
\vspace*{-3mm}
\centerline{\hspace*{3mm}\psfig{file=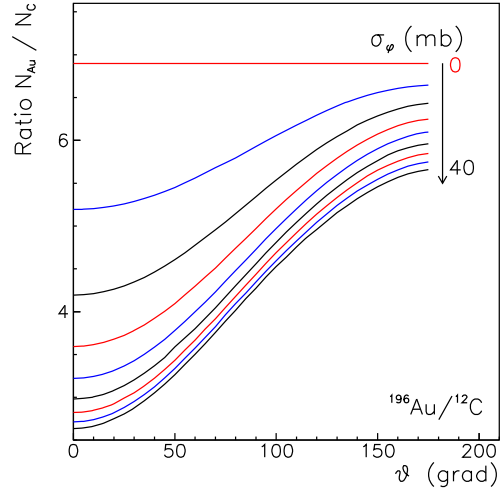,width=7.5cm}}
\vspace*{-3mm}
\caption{The ratio of the effective collision numbers calculated for $^{196}Au$
and $^{12}C$ targets as a function of $\phi$-meson production angle shown for
the different $\sigma_\phi$ within the range from 0 to 40 mb with a step size of
5 mb.}
\label{cofi4b}
\end{figure}

The calculations were done for the $^{12}C$, $^{64}Cu$, $^{108}Ag$ and
$^{196}Au$ nuclear targets, which is in line with  the targets
proposed~\cite{Hartmann} for the measurements at COSY.
Figs.~\ref{cofi4a}-\ref{cofi4b} shows the calculated ratios as a function of the
$\phi$-meson production angle. The lines indicate the results for different
$\sigma_\phi$ given within the range from 0 to 40 mb with a step size of 5 mb. 

The calculations indicate substantial angular dependence of the ratio. As we
showed previously, the analysis of the $A$-dependence for the forward particle
production is most preferable for several reasons. At forward angles, the
reaction mechanism can be formulated within an eikonal basis and contains less
theoretical uncertainties than the quasi-classical approximation. Furthermore,
due to the scattering dynamics any possible multiple processes contribute less
to the forward particle production.
 
Moreover, at forward angles the ratios indicate reasonable sensitivity to the
distortion cross section when $\sigma_\phi$ stands below $\simeq$20~MeV. At
large angles and  for large $\sigma_\phi$ the analysis requires very high
precision data.

For completeness let us to illustrate how to use the figures with the calculated
ratios. We take as an example the ratio of the effective collision numbers from
$^{108}Ag$ and $^{12}C$ nuclei shown in Fig.~\ref{cofi4}. Let us consider the
production at forward angles, {\it i. e.} $\theta{\le}10^o$. If there is no
distortion of the incident proton and final $\phi$-meson the ratio equals that
given by the target mass numbers leading to $R$=9. Due to the distortion of the
incident proton and neglecting the distortion of the produced $\phi$-meson one
finds that $R$=4.6. That is the maximum value given by the direct $\phi$-meson
production mechanism.  The $A^{2/3}$ dependence corresponds to $R$=4.3 and leads
to a $\phi$-meson distortion cross section of less than 5 mb. The $A^{1/3}$
dependence results in a  ratio of ${\simeq}$2.1 and corresponds to
$\sigma_\phi{>}$40 mb. 

Now the question arises if the measured ratio is larger than $R$=4.6. This
explicitly indicates the contribution from multiple or two-step processes, as is
illustrated by Fig.~\ref{cofi10a}. In that case the extraction of the
$\phi$-meson distortion in nuclear matter is much more model-dependent. Then one
could use additional kinematical constraints in order to isolate direct
production mechanism, as was done for instance in Ref.~\cite{Barmin}.

Finally, we can estimate the $A$-dependence in case the distortion cross section
is $\sigma_\phi$=10 mb. Then one might expect at forward angles the ratio for
$\phi$-meson production from $^{108}Ag$ and $^{12}C$ targets to be $R$=3.2 as is
illustrated by Fig.~\ref{cofi4}. This corresponds to a mass dependence of
${\propto}A^{0.52}$. This result is compatible with in-medium width of
$\simeq$30 MeV for an average $\phi$-meson momentum of 500 MeV/c. 

\section{Conclusions}

A systematic analysis of  the $A$-dependence of $\phi$-meson production in
proton-nucleus collisions has been carried out. We discuss the application of an
eikonal formalism, corrections due to multiple scattering and the extention to
large angle production processes. Furthermore, the $A$-dependence due to
two-step production mechanisms and multi-step processes are investigated in
detail. We provide all formulas frequently used in the analysis of nuclear data
and study their compatibility and conditions of applicability.

The $\phi$-meson distortion cross section $\sigma_\phi$ was evaluated from the
available nuclear data. It was found that different measurements result in
different values of $\sigma_\phi$ ranging from 0 to 14 mb. Unfortunately, at
present the uncertainties of the experimental results are too large to draw
definite conclusions.

We also discuss an interpretation of the $\phi$-meson distortion in nuclear
matter and give the relation between various frequently used variables, such as
in-medium width, distortion cross section and scattering length. Furthermore, we
show the estimates for $\sigma_\phi$ in the vacuum obtained by VDM, the additive
quark model, QCD sum rules and other theoretical frameworks available. Moreover,
we collect  predictions for the $\phi$-meson modification in matter. While most
of the estimates are in reasonable agreement with $\sigma_\phi{\simeq}$10 mb at
$\phi$-meson energies above 3 GeV there are very large uncertainties at lower
energies.

To resolve the unsatisfactory current situation, we propose to study the
$A$-dependence of $\phi$-meson production from $p{+}A$ collisions at COSY
energies.  We provide detailed calculations of the ratios of $\phi$-meson
production cross sections from different nuclear targets. Our results can
directly be used for the evaluation of the $\sigma_\phi$ from such 
measurements.

\begin{acknowledgement}
We thank M.~Hartmann, F.~Huang, Yu.~Kiselev, U.~Mosel, E.~Oset, E.~Paryev,
J.~Tjon and K.~Tsu\-shi\-ma for discussions. This work was partially  supported 
by the Helmholtz Association (Virtual institute ``Spin and strong QCD'',
VH-VI-231) and the Deutsche Forschungsgemeinschaft  through funds provided to
the SFB/TR 16 ``Subnuclear Structure of Matter'' and grant DFG 436 RUS
113/924/0-1. This research is part of the \, EU Integrated \, Infrastructure \,
Initiative Hadron Physics Project under contract  number \, RII3-CT-2004-506078.
\, A.S. acknowledges support from the JLab grant \, SURA-06-C0452  and the  \,
COSY FFE grant No. 41760632 (COSY-085).  
\end{acknowledgement}

\end{document}